%% file: hadron2011.tex
\documentclass[a4paper,11pt]{article}

\usepackage{contribution}

\input{econfmacros}
\input{contributionmacros}

\begin{document}

\input{contribution}

\end{document}

%% file: econfmacros.tex
\newcommand{\weblink}[2][]{%
    \ifthenelse{\equal{#1}{}}%
    {\textnormal{\url{#2}}}%
    {\textnormal{\href{#2}{#1}}}%
}

\def\beq{\begin{equation}}
\def\eeq#1{\label{#1}\end{equation}}
\def\eeqn{\end{equation}}

\def\beqa{\begin{eqnarray}}
\def\eeqa#1{\label{#1}\end{eqnarray}}
\def\eeqan{\end{eqnarray}}

\def\st{\scriptstyle}

\let\bar=\overbar

\def\eg{{\it e.g.}}

\def\D{{\cal D}}
\def\L{{\cal L}}

\def\W{{\cal W}}

\def\Dslash{\not{\hbox{\kern-4pt $D$}}}
\def\dslash{\not{\hbox{\kern-2pt $\del$}}}

\def\BR{\mbox{\rm BR}}
\def\ee{e^+e^-}

\def\msb{{\bar{\ssstyle M \kern -1pt S}}}

\def\eps{\epsilon}

\def\s#1{\widetilde{#1}}

%% file: contributionmacros.tex
\newcommand{\contribution}[7][]{%
  \clearpage
  \thispagestyle{plain}
  \ifthenelse{\equal{#1}{}}
  {\hypersetup{pdftitle={#2}}}
  {\hypersetup{pdftitle={#1}}}
  \hypersetup{pdfauthor={{#3} {#4}}}
  {\centering\normalfont\LARGE\bfseries\sffamily #2 \par\nobreak}
  \lhead{}
  \chead{%
    \textit{\footnotesize XIV International Conference on Hadron Spectroscopy
      (\weblink[\textit{hadron2011}]{http://www.hadron2011.de}), 13-17 June 2011, Munich, Germany}%
  }
  \rhead{}
  \bigskip
  \begin{center}
    {#3} {#4}\ifthenelse{\equal{#6}{}}{}{\footnote{\weblink[#6]{mailto:#6}}}
    \ifthenelse{\equal{#7}{}}{}{#7} \\
    \textit{#5}
  \end{center}
  \bigskip
}

\renewcommand{\abstract}[1]{%
  \begin{center}
    \begin{minipage}{0.85\textwidth}
      \begin{footnotesize}
        #1
      \end{footnotesize}
    \end{minipage}
  \end{center}
  \bigskip
}

%% file: contribution.tex
%
%
%
%
%
{  


\def\hbabar{\mbox{{\huge\bf\sl B}\hspace{-0.1em}{\LARGE\bf\sl A}\hspace{-0.03em}{\huge\bf\sl B}\hspace{-0.1em}{\LARGE\bf\sl A\hspace{-0.03em}R}}}
\def\Lbabar{\mbox{{\LARGE\sl B}\hspace{-0.15em}{\Large\sl A}\hspace{-0.07em}{\LARGE\sl B}\hspace{-0.15em}{\Large\sl A\hspace{-0.02em}R}}}
\def\lbabar{\mbox{{\large\sl B}\hspace{-0.4em} {\normalsize\sl A}\hspace{-0.03em}{\large\sl B}\hspace{-0.4em} {\normalsize\sl A\hspace{-0.02em}R}}}
\def\babar{\mbox{\slshape B\kern-0.1em{\smaller A}\kern-0.1em
    B\kern-0.1em{\smaller A\kern-0.2em R}}}

\def\belle{Belle}

\let\emi\en
\def\electron   {\ensuremath{e}\xspace}
\def\en         {\ensuremath{e^-}\xspace}   
\def\ep         {\ensuremath{e^+}\xspace}
\def\epm        {\ensuremath{e^\pm}\xspace} 
\def\epem       {\ensuremath{e^+e^-}\xspace}
\def\ee         {\ensuremath{e^-e^-}\xspace}

\def\mmu        {\ensuremath{\mu}\xspace}
\def\mup        {\ensuremath{\mu^+}\xspace}
\def\mun        {\ensuremath{\mu^-}\xspace} 
\def\mumu       {\ensuremath{\mu^+\mu^-}\xspace}
\def\mtau       {\ensuremath{\tau}\xspace}

\def\taup       {\ensuremath{\tau^+}\xspace}
\def\taum       {\ensuremath{\tau^-}\xspace}
\def\tautau     {\ensuremath{\tau^+\tau^-}\xspace}

\def\ellm       {\ensuremath{\ell^-}\xspace}
\def\ellp       {\ensuremath{\ell^+}\xspace}
\def\ellell     {\ensuremath{\ell^+ \ell^-}\xspace}

\def\nub        {\ensuremath{\overline{\nu}}\xspace}
\def\nunub      {\ensuremath{\nu{\overline{\nu}}}\xspace}
\def\nub        {\ensuremath{\overline{\nu}}\xspace}
\def\nunub      {\ensuremath{\nu{\overline{\nu}}}\xspace}
\def\nue        {\ensuremath{\nu_e}\xspace}
\def\nueb       {\ensuremath{\nub_e}\xspace}
\def\nuenueb    {\ensuremath{\nue\nueb}\xspace}
\def\num        {\ensuremath{\nu_\mu}\xspace}
\def\numb       {\ensuremath{\nub_\mu}\xspace}
\def\numnumb    {\ensuremath{\num\numb}\xspace}
\def\nut        {\ensuremath{\nu_\tau}\xspace}
\def\nutb       {\ensuremath{\nub_\tau}\xspace}
\def\nutnutb    {\ensuremath{\nut\nutb}\xspace}
\def\nul        {\ensuremath{\nu_\ell}\xspace}
\def\nulb       {\ensuremath{\nub_\ell}\xspace}
\def\nulnulb    {\ensuremath{\nul\nulb}\xspace}


\def\g     {\ensuremath{\gamma}\xspace}
\def\gaga  {\ensuremath{\gamma\gamma}\xspace}  
\def\ggstar{\ensuremath{\gamma\gamma^*}\xspace}

\def\ega    {\ensuremath{e\gamma}\xspace}
\def\game   {\ensuremath{\gamma e^-}\xspace}
\def\epemg  {\ensuremath{e^+e^-\gamma}\xspace}


\def\H      {\ensuremath{H^0}\xspace}
\def\Hp     {\ensuremath{H^+}\xspace}
\def\Hm     {\ensuremath{H^-}\xspace}
\def\Hpm    {\ensuremath{H^\pm}\xspace}
\def\W      {\ensuremath{W}\xspace}
\def\Wp     {\ensuremath{W^+}\xspace}
\def\Wm     {\ensuremath{W^-}\xspace}
\def\Wpm    {\ensuremath{W^\pm}\xspace}
\def\Z      {\ensuremath{Z^0}\xspace}


\def\q     {\ensuremath{q}\xspace}
\def\qbar  {\ensuremath{\overline q}\xspace}
\def\qqbar {\ensuremath{q\overline q}\xspace}
\def\u     {\ensuremath{u}\xspace}
\def\ubar  {\ensuremath{\overline u}\xspace}
\def\uubar {\ensuremath{u\overline u}\xspace}
\def\d     {\ensuremath{d}\xspace}
\def\dbar  {\ensuremath{\overline d}\xspace}
\def\ddbar {\ensuremath{d\overline d}\xspace}
\def\s     {\ensuremath{s}\xspace}
\def\sbar  {\ensuremath{\overline s}\xspace}
\def\ssbar {\ensuremath{s\overline s}\xspace}
\def\c     {\ensuremath{c}\xspace}
\def\cbar  {\ensuremath{\overline c}\xspace}
\def\ccbar {\ensuremath{c\overline c}\xspace}
\def\b     {\ensuremath{b}\xspace}
\def\bbar  {\ensuremath{\overline b}\xspace}
\def\bbbar {\ensuremath{b\overline b}\xspace}
\def\t     {\ensuremath{t}\xspace}
\def\tbar  {\ensuremath{\overline t}\xspace}
\def\tbar  {\ensuremath{\overline t}\xspace}
\def\ttbar {\ensuremath{t\overline t}\xspace}
\def\ccbars {\ensuremath{c\cbar\s}\xspace}


\def\piz   {\ensuremath{\pi^0}\xspace}
\def\pizs  {\ensuremath{\pi^0\mbox\,\rm{s}}\xspace}
\def\ppz   {\ensuremath{\pi^0\pi^0}\xspace}
\def\pip   {\ensuremath{\pi^+}\xspace}
\def\pim   {\ensuremath{\pi^-}\xspace}
\def\pipi  {\ensuremath{\pi^+\pi^-}\xspace}
\def\pipm  {\ensuremath{\pi^\pm}\xspace}
\def\pimp  {\ensuremath{\pi^\mp}\xspace}

\def\kaon  {\ensuremath{K}\xspace}
\def\Kbar  {\kern 0.2em\overline{\kern -0.2em K}{}\xspace}
\def\Kb    {\ensuremath{\Kbar}\xspace}
\def\Kz    {\ensuremath{K^0}\xspace}
\def\Kzb   {\ensuremath{\Kbar^0}\xspace}
\def\KzKzb {\ensuremath{\Kz \kern -0.16em \Kzb}\xspace}
\def\Kp    {\ensuremath{K^+}\xspace}
\def\Km    {\ensuremath{K^-}\xspace}
\def\Kpm   {\ensuremath{K^\pm}\xspace}
\def\Kmp   {\ensuremath{K^\mp}\xspace}
\def\KpKm  {\ensuremath{\Kp \kern -0.16em \Km}\xspace}
\def\KS    {\ensuremath{K^0_{\scriptscriptstyle S}}\xspace} 
\def\KL    {\ensuremath{K^0_{\scriptscriptstyle L}}\xspace} 
\def\Kstarz  {\ensuremath{K^{*0}}\xspace}
\def\Kstarzb {\ensuremath{\Kbar^{*0}}\xspace}
\def\Kstar   {\ensuremath{K^*}\xspace}
\def\Kstarb  {\ensuremath{\Kbar^*}\xspace}
\def\Kstarp  {\ensuremath{K^{*+}}\xspace}
\def\Kstarm  {\ensuremath{K^{*-}}\xspace}
\def\Kstarpm {\ensuremath{K^{*\pm}}\xspace}
\def\Kstarmp {\ensuremath{K^{*\mp}}\xspace}

\newcommand{\etapr}{\ensuremath{\eta^{\prime}}\xspace}


\def\azero   {\ensuremath{a_0}\xspace}
\def\azeroz  {\ensuremath{a_0^0}\xspace}
\def\azerop  {\ensuremath{a_0^+}\xspace}
\def\azerom  {\ensuremath{a_0^-}\xspace}
\def\azeropm {\ensuremath{a_0^{\pm}}\xspace}

\def\aone    {\ensuremath{a_1(1260)}\xspace}
\def\aonez   {\ensuremath{a_1^0(1260)}\xspace}
\def\aonep   {\ensuremath{a_1^+(1260)}\xspace}
\def\aonem   {\ensuremath{a_1^-(1260)}\xspace}
\def\aonepm  {\ensuremath{a_1^{\pm}(1260)}\xspace}

\def\bone    {\ensuremath{b_1}\xspace}
\def\bonez   {\ensuremath{b_1^0}\xspace}
\def\bonep   {\ensuremath{b_1^+}\xspace}
\def\bonem   {\ensuremath{b_1^-}\xspace}
\def\bonepm  {\ensuremath{b_1^{\pm}}\xspace}

\def\rhoz    {\ensuremath{\rho^0}\xspace}
\def\rhop    {\ensuremath{\rho^+}\xspace}
\def\rhom    {\ensuremath{\rho^-}\xspace}
\def\rhopm   {\ensuremath{\rho^{\pm}}\xspace}
\def\rhomp   {\ensuremath{\rho^{\mp}}\xspace}

\def\fz      {\ensuremath{f_0(980)}\xspace}


\def\D       {\ensuremath{D}\xspace}
\def\Dbar    {\kern 0.2em\overline{\kern -0.2em D}{}\xspace}
\def\Db      {\ensuremath{\Dbar}\xspace}
\def\Dz      {\ensuremath{D^0}\xspace}
\def\Dzb     {\ensuremath{\Dbar^0}\xspace}
\def\DzDzb   {\ensuremath{\Dz {\kern -0.16em \Dzb}}\xspace}
\def\Dp      {\ensuremath{D^+}\xspace}
\def\Dm      {\ensuremath{D^-}\xspace}
\def\Dpm     {\ensuremath{D^\pm}\xspace}
\def\Dmp     {\ensuremath{D^\mp}\xspace}
\def\DpDm    {\ensuremath{\Dp {\kern -0.16em \Dm}}\xspace}
\def\Dstar   {\ensuremath{D^*}\xspace}
\def\Dstarb  {\ensuremath{\Dbar^*}\xspace}
\def\Dstarz  {\ensuremath{D^{*0}}\xspace}
\def\Dstarzb {\ensuremath{\Dbar^{*0}}\xspace}
\def\Dstarp  {\ensuremath{D^{*+}}\xspace}
\def\Dstarm  {\ensuremath{D^{*-}}\xspace}
\def\Dstarpm {\ensuremath{D^{*\pm}}\xspace}
\def\Dstarmp {\ensuremath{D^{*\mp}}\xspace}
\def\Ds      {\ensuremath{D^+_s}\xspace}
\def\Dsb     {\ensuremath{\Dbar^+_s}\xspace}
\def\Dss     {\ensuremath{D^{*+}_s}\xspace}
\def\Dmix    {\ensuremath{\Dz-\Dzb}\xspace}

\newcommand{\dstr}{\ensuremath{\Dstar}\xspace}
\newcommand{\dstrstr}{\ensuremath{D^{**}}\xspace}
\newcommand{\dsp}{\ensuremath{\Dstarp}\xspace}
\newcommand{\dsm}{\ensuremath{\Dstarm}\xspace}
\newcommand{\dsz}{\ensuremath{\Dstarz}\xspace}

\def\B       {\ensuremath{B}\xspace}
\def\Bbar    {\kern 0.18em\overline{\kern -0.18em B}{}\xspace}
\def\Bb      {\ensuremath{\Bbar}\xspace}
\def\BB      {\ensuremath{B\Bbar}\xspace} 
\def\Bz      {\ensuremath{B^0}\xspace}
\def\Bzb     {\ensuremath{\Bbar^0}\xspace}
\def\BzBzb   {\ensuremath{\Bz {\kern -0.16em \Bzb}}\xspace}
\def\Bu      {\ensuremath{B^+}\xspace}
\def\Bub     {\ensuremath{B^-}\xspace}
\def\Bp      {\ensuremath{\Bu}\xspace}
\def\Bm      {\ensuremath{\Bub}\xspace}
\def\Bpm     {\ensuremath{B^\pm}\xspace}
\def\Bmp     {\ensuremath{B^\mp}\xspace}
\def\BpBm    {\ensuremath{\Bu {\kern -0.16em \Bub}}\xspace}
\def\Bs      {\ensuremath{B_s}\xspace}
\def\Bsb     {\ensuremath{\Bbar_s}\xspace}
\def\Btag    {\ensuremath{\B_{\mathrm{tag}}}\xspace}
\def\Brec    {\ensuremath{\B_{\mathrm{rec}}}\xspace}
\def\Bflav   {\ensuremath{\B_{\mathrm{flav}}}\xspace}
\def\Bmix    {\ensuremath{\Bz-\Bzb}\xspace}

\def\BorBbar    {\kern 0.18em\optbar{\kern -0.18em B}{}\xspace}
\def\DorDbar    {\kern 0.18em\optbar{\kern -0.18em D}{}\xspace}
\def\KorKbar    {\kern 0.18em\optbar{\kern -0.18em K}{}\xspace}


\def\jpsi     {\ensuremath{{J\mskip -3mu/\mskip -2mu\psi\mskip 2mu}}\xspace}
\def\psitwos  {\ensuremath{\psi{(2S)}}\xspace}
\def\psiprpr  {\ensuremath{\psi(3770)}\xspace}
\def\etac     {\ensuremath{\eta_c}\xspace}
\def\chiczero {\ensuremath{\chi_{c0}}\xspace}
\def\chicone  {\ensuremath{\chi_{c1}}\xspace}
\def\chictwo  {\ensuremath{\chi_{c2}}\xspace}
\mathchardef\Upsilon="7107
\def\Y#1S{\ensuremath{\Upsilon{(#1S)}}\xspace}
\def\OneS  {\Y1S}
\def\TwoS  {\Y2S}
\def\ThreeS {\Y3S}
\def\FourS {\Y4S}
\def\FiveS {\Y5S}

\def\chic#1{\ensuremath{\chi_{c#1}}\xspace} 


\def\proton      {\ensuremath{p}\xspace}
\def\antiproton  {\ensuremath{\overline p}\xspace}
\def\neutron     {\ensuremath{n}\xspace}
\def\antineutron {\ensuremath{\overline n}\xspace}

\mathchardef\Deltares="7101
\mathchardef\Xi="7104
\mathchardef\Lambda="7103
\mathchardef\Sigma="7106
\mathchardef\Omega="710A

\def\Deltabar {\kern 0.25em\overline{\kern -0.25em \Deltares}{}\xspace}
\def\Lbar     {\kern 0.2em\overline{\kern -0.2em\Lambda\kern 0.05em}\kern-0.05em{}\xspace}
\def\Sigbar   {\kern 0.2em\overline{\kern -0.2em \Sigma}{}\xspace}
\def\Xibar    {\kern 0.2em\overline{\kern -0.2em \Xi}{}\xspace}
\def\Obar     {\kern 0.2em\overline{\kern -0.2em \Omega}{}\xspace}
\def\Nbar     {\kern 0.2em\overline{\kern -0.2em N}{}\xspace}
\def\Xb       {\kern 0.2em\overline{\kern -0.2em X}{}\xspace}

\def\X {\ensuremath{X}\xspace}


\def\BR         {{\ensuremath{\cal B}\xspace}}
\def\BRtauptoe  {\ensuremath{\BR(\taup \to \ep)}\xspace}
\def\BRtaumtoe  {\ensuremath{\BR(\taum \to \en)}\xspace}
\def\BRtauptomu {\ensuremath{\BR(\taup \to \mup)}\xspace}
\def\BRtaumtomu {\ensuremath{\BR(\taum \to \mun)}\xspace}


\newcommand{\etaprepp}{\ensuremath{\etapr \to \eta \pipi}\xspace}
\newcommand{\etaprrg} {\ensuremath{\etapr \to \rho^0 \g}\xspace}


\def\bpsiks     {\ensuremath{\Bz \to \jpsi \KS}\xspace}
\def\bpsikst    {\ensuremath{\Bz \to \jpsi \Kstar}\xspace}
\def\bpsikl     {\ensuremath{\Bz \to \jpsi \KL}\xspace}
\def\bpsikzeropi{\ensuremath{\Bz \to \jpsi \Kstarz (\to \KL \piz)}\xspace}
\def\bpsikpluspi{\ensuremath{\Bu \to \jpsi \Kstarp (\to \KL \pip)}\xspace}
\def\bpsikpi    {\ensuremath{\Bz/\Bzb \to \jpsi (\to \mumu) \Kpm \pimp}\xspace}
\def\bupsik     {\ensuremath{\Bpm \to \jpsi (\to \mumu) \Kpm}\xspace}
\def\bpsiX      {\ensuremath{\Bz \to \jpsi \X}\xspace}

\def\Bzbtomu    {\ensuremath{\Bzb \to \mu \X}\xspace}
\def\Bzbtox     {\ensuremath{\Bzb \to \X}\xspace}
\def\Bztopipi   {\ensuremath{\Bz \to \pipi}\xspace}
\def\Bztokpi    {\ensuremath{\Bz \to \Kpm \pimp}\xspace}
\def\Bztorhopi  {\ensuremath{\Bz \to \rho^+ \pim}\xspace}
\def\Bztorhorho {\ensuremath{\Bz \to \rho \rho}\xspace}
\def\Bztokrho   {\ensuremath{\Bz \to K \rho}\xspace}
\def\Bztokstpi  {\ensuremath{\Bz \to \Kstar \pi}\xspace}
\def\Bztoapi    {\ensuremath{\Bz \to a_1 \pi}\xspace}
\def\Bztodd     {\ensuremath{\Bz \to \DpDm}\xspace}
\def\Bztodstd   {\ensuremath{\Bz \to \Dstarp \Dm}\xspace}
\def\Bztodstdst {\ensuremath{\Bz \to \Dstarp \Dstarm}\xspace}

\def\BtoDK      {\ensuremath{B \to DK}\xspace}
\def\Btodstlnu  {\ensuremath{B \to \Dstar \ell \nu}\xspace}
\def\Btodstdlnu {\ensuremath{B \to \Dstar(D) \ell \nu}\xspace}
\def\Btorholnu  {\ensuremath{B \to \rho \ell \nu}\xspace}
\def\Btopilnu   {\ensuremath{B \to \pi \ell \nu}\xspace}

\def\Btoetah    {\ensuremath{B \to \eta h}\xspace}
\def\Btoetaph   {\ensuremath{B \to \etapr h}\xspace}

\newcommand{\Betaprks}{\ensuremath{\Bz \to \etapr \KS}\xspace}
\newcommand{\Betaprkz}{\ensuremath{\Bz \to \etapr \Kz}\xspace}

\def\btosgam    {\ensuremath{b \to s \g}\xspace}
\def\btodgam    {\ensuremath{b \to d \g}\xspace}
\def\btosll     {\ensuremath{b \to s \ellell}\xspace}
\def\btosnunu   {\ensuremath{b \to s \nunub}\xspace}
\def\btosgaga   {\ensuremath{b \to s \gaga}\xspace}
\def\btosglue   {\ensuremath{b \to s g}\xspace}


\def\upsbb   {\ensuremath{\FourS \to \BB}\xspace}
\def\upsbzbz {\ensuremath{\FourS \to \BzBzb}\xspace}
\def\upsbpbm {\ensuremath{\FourS \to \BpBm}\xspace}
\def\upspsikl{\ensuremath{\FourS \to (\bpsikl) (\Bzbtox)}\xspace}


\def\tauptoe    {\ensuremath{\taup \to \ep \nunub}\xspace}
\def\taumtoe    {\ensuremath{\taum \to \en \nunub}\xspace}
\def\tauptomu   {\ensuremath{\taup \to \mup \nunub}\xspace}
\def\taumtomu   {\ensuremath{\taum \to \mun \nunub}\xspace}
\def\tauptopi   {\ensuremath{\taup \to \pip \nub}\xspace}
\def\taumtopi   {\ensuremath{\taum \to \pim \nu}\xspace}


\def\ggtopi     {\ensuremath{\gaga \to \pipi}\xspace}
\def\ggtopiz    {\ensuremath{\gaga \to \ppz}\xspace}
\def\ggstox     {\ensuremath{\ggstar \to \X(1420) \to \kaon \kaon \pi}\xspace}
\def\ggstoeta   {\ensuremath{\ggstar \to \eta(550) \to \pipi \piz}\xspace}


\def\ptot       {\mbox{$p$}\xspace}
\def\pxy        {\mbox{$p_T$}\xspace}
\def\pt         {\mbox{$p_T$}\xspace}
\def\mes        {\mbox{$m_{\rm ES}$}\xspace}
\def\mec        {\mbox{$m_{\rm EC}$}\xspace}
\def\DeltaE     {\mbox{$\Delta E$}\xspace}

\def\pbcm {\ensuremath{p^*_{\Bz}}\xspace}


\def\mphi       {\mbox{$\phi$}\xspace}
\def\mtheta     {\mbox{$\theta$}\xspace}
\def\ctheta     {\mbox{$\cos\theta$}\xspace}


\newcommand{\tev}{\ensuremath{\mathrm{\,Te\kern -0.1em V}}\xspace}
\newcommand{\gev}{\ensuremath{\mathrm{\,Ge\kern -0.1em V}}\xspace}
\newcommand{\mev}{\ensuremath{\mathrm{\,Me\kern -0.1em V}}\xspace}
\newcommand{\kev}{\ensuremath{\mathrm{\,ke\kern -0.1em V}}\xspace}
\newcommand{\ev}{\ensuremath{\mathrm{\,e\kern -0.1em V}}\xspace}
\newcommand{\gevc}{\ensuremath{{\mathrm{\,Ge\kern -0.1em V\!/}c}}\xspace}
\newcommand{\mevc}{\ensuremath{{\mathrm{\,Me\kern -0.1em V\!/}c}}\xspace}
\newcommand{\gevcc}{\ensuremath{{\mathrm{\,Ge\kern -0.1em V\!/}c^2}}\xspace}
\newcommand{\mevcc}{\ensuremath{{\mathrm{\,Me\kern -0.1em V\!/}c^2}}\xspace}


\def\syin {\ensuremath{^{\prime\prime}}\xspace}
\def\inch   {\ensuremath{\rm \,in}\xspace} 
\def\ft   {\ensuremath{\rm \,ft}\xspace}
\def\km   {\ensuremath{{\rm \,km}}\xspace}
\def\m    {\ensuremath{{\rm \,m}}\xspace}
\def\cm   {\ensuremath{{\rm \,cm}}\xspace}
\def\mm   {\ensuremath{{\rm \,mm}}\xspace}
\def\mum  {\ensuremath{{\,\mu\rm m}}\xspace}
\def\nm   {\ensuremath{{\rm \,nm}}\xspace}
\def\fm   {\ensuremath{{\rm \,fm}}\xspace}
\def\nm         {\ensuremath{{\rm \,nm}}\xspace}   

\def\cma  {\ensuremath{{\rm \,cm}^2}\xspace}
\def\mma  {\ensuremath{{\rm \,mm}^2}\xspace}
\def\muma {\ensuremath{{\,\mu\rm m}^2}\xspace}

\def\barn      {\ensuremath{{\rm \,b}}\xspace}
\def\barnhyph  {\ensuremath{{\rm -b}}\xspace}
\def\mbarn     {\ensuremath{{\rm \,mb}}\xspace}
\def\mbarnhyph {\ensuremath{{\rm -mb}}\xspace}
\def\nb        {\ensuremath{{\rm \,nb}}\xspace}
\def\pb {\ensuremath{{\rm \,pb}}\xspace}
\def\fb   {\ensuremath{\mbox{\,fb}}\xspace}

\def\invnb {\ensuremath{\mbox{\,nb}^{-1}}\xspace}
\def\invpb {\ensuremath{\mbox{\,pb}^{-1}}\xspace}
\def\invfb   {\ensuremath{\mbox{\,fb}^{-1}}\xspace}
\def\invab   {\ensuremath{\mbox{\,ab}^{-1}}\xspace}


\def\gm   {\ensuremath{\rm \,g}\xspace}

\def\sec  {\ensuremath{\rm {\,s}}\xspace}       
\def\ms   {\ensuremath{{\rm \,ms}}\xspace}     
\def\mus  {\ensuremath{\,\mu{\rm s}}\xspace}    
\def\ns   {\ensuremath{{\rm \,ns}}\xspace}      
\def\ps   {\ensuremath{{\rm \,ps}}\xspace}  
\def\fs   {\ensuremath{\rm \,fs}\xspace}




\def\eg {e.g.}
\def\etc {etc.}

\def\Xrad {\ensuremath{X_0}\xspace}
\def\NIL  {\ensuremath{\lambda_{int}}\xspace}
\let\dgr\degrees
\def\todo        {{\it{\red To\:be\:completed.}}}

\def\cms  {\ensuremath{{\rm \,cm}^{-2} {\rm s}^{-1}}\xspace}

\def\mic  {\ensuremath{\,\mu{\rm C}}\xspace}
\def\krad {\ensuremath{\rm \,krad}\xspace}
\def\cmc  {\ensuremath{{\rm \,cm}^3}\xspace}
\def\yr   {\ensuremath{\rm \,yr}\xspace}
\def\hr   {\ensuremath{\rm \,hr}\xspace}
\def\degc {\ensuremath{^\circ}{C}\xspace}
\def\degk {\ensuremath {\rm K}\xspace}
\def\degrees {\ensuremath{^{\circ}}\xspace}
\def\mrad {\ensuremath{\rm \,mrad}\xspace}               
\def\rad{ \ensuremath{\rm \,rad}\xspace}
\def\mradhyph {\ensuremath{\rm -mr}\xspace}
\def\sx    {\ensuremath{\sigma_x}\xspace}    
\def\sy    {\ensuremath{\sigma_y}\xspace}   
\def\sz    {\ensuremath{\sigma_z}\xspace}    


\def\order {{\ensuremath{\cal O}}\xspace}
\def\L     {{\ensuremath{\cal L}}\xspace}
\def\calL  {{\ensuremath{\cal L}}\xspace}
\def\calS  {{\ensuremath{\cal S}}\xspace}
\def\calA  {{\ensuremath{\cal A}}\xspace}
\def\calD  {{\ensuremath{\cal D}}\xspace}
\def\calR  {{\ensuremath{\cal R}}\xspace}
\def\calB  {{\ensuremath{\cal B}}\xspace}

\def\ra                 {\ensuremath{\rightarrow}\xspace}
\def\to                 {\ensuremath{\rightarrow}\xspace}

\newcommand{\stat}{\ensuremath{\mathrm{(stat)}}\xspace}
\newcommand{\syst}{\ensuremath{\mathrm{(syst)}}\xspace}

\def\pep2{PEP-II}
\def\BF{$B$ Factory}
\def\abf {asymmetric \BF}

\newcommand{\inverse} {\ensuremath{^{-1}}\xspace}
\newcommand{\dedx}    {\ensuremath{\mathrm{d}\hspace{-0.1em}E/\mathrm{d}x}\xspace}
\newcommand{\chisq}   {\ensuremath{\chi^2}\xspace}
\newcommand{\delm}    {\ensuremath{m_{\dstr}-m_{\dz}}\xspace}
\newcommand{\lum}     {\ensuremath{\mathcal{L}}\xspace}

\def\gsim{{~\raise.15em\hbox{$>$}\kern-.85em
          \lower.35em\hbox{$\sim$}~}\xspace}
\def\lsim{{~\raise.15em\hbox{$<$}\kern-.85em
          \lower.35em\hbox{$\sim$}~}\xspace}

\def\qsq                {\ensuremath{q^2}\xspace}

\def\kbytes     {\ensuremath{{\rm \,kbytes}}\xspace}
\def\kbsps      {\ensuremath{{\rm \,kbytes/s}}\xspace}
\def\kbits      {\ensuremath{{\rm \,kbits}}\xspace}
\def\kbitss     {\ensuremath{{\rm \,kbits/s}}\xspace}
\def\mbsps      {\ensuremath{{\rm \,Mbits/s}}\xspace}
\def\mbytes     {\ensuremath{{\rm \,Mbytes}}\xspace}
\def\mbps       {\ensuremath{{\rm \,Mbyte/s}}\xspace}
\def\mbsps      {\ensuremath{{\rm \,Mbytes/s}}\xspace}
\def\gbsps      {\ensuremath{{\rm \,Gbits/s}}\xspace}
\def\gbytes     {\ensuremath{{\rm \,Gbytes}}\xspace}
\def\gbsps      {\ensuremath{{\rm \,Gbytes/s}}\xspace}
\def\tbytes     {\ensuremath{{\rm \,Tbytes}}\xspace}
\def\tbpy       {\ensuremath{{\rm \,Tbytes/yr}}\xspace}

\def\kHz        {\ensuremath{{\rm \,kHz}}\xspace}
\def\MHz        {\ensuremath{{\rm \,MHz}}\xspace}

\def\Watt       {\ensuremath{{\rm \,W}}\xspace}
\def\miWatt     {\ensuremath{{\rm \,mW}}\xspace}
\def\muWatt     {\ensuremath{\,\mu{\rm W}}\xspace} 
\def\MWatt     {\ensuremath{\,{\rm MW}}\xspace} 


\newcommand{\as}{\ensuremath{\alpha_{\scriptscriptstyle S}}\xspace}
\newcommand{\MSb}{\ensuremath{\overline{\mathrm{MS}}}\xspace}
\newcommand{\LMSb}{%
  \ensuremath{\Lambda_{\overline{\scriptscriptstyle\mathrm{MS}}}}\xspace
}


\newcommand{\tw}{\ensuremath{\theta_{\scriptscriptstyle W}}\xspace}
\newcommand{\twb}{%
  \ensuremath{\overline{\theta}_{\scriptscriptstyle W}}\xspace
}
\newcommand{\Afb}[1]{{\ensuremath{A_{\scriptscriptstyle FB}^{#1}}}\xspace}
\newcommand{\gv}[1]{{\ensuremath{g_{\scriptscriptstyle V}^{#1}}}\xspace}
\newcommand{\ga}[1]{{\ensuremath{g_{\scriptscriptstyle A}^{#1}}}\xspace}
\newcommand{\gvb}[1]{{\ensuremath{\overline{g}_{\scriptscriptstyle V}^{#1}}}\xspace}
\newcommand{\gab}[1]{{\ensuremath{\overline{g}_{\scriptscriptstyle A}^{#1}}}\xspace}


\def\eps  {\varepsilon\xspace}
\def\epsK {\varepsilon_K\xspace}
\def\epsB {\varepsilon_B\xspace}
\def\epsp {\varepsilon^\prime_K\xspace}

\def\CP      {\ensuremath{C\!P}\xspace}
\def\CPT     {\ensuremath{C\!PT}\xspace}
\def\C       {\ensuremath{C}\xspace}
\def\P       {\ensuremath{P}\xspace}
\def\T       {\ensuremath{T}\xspace}

\def\rhobar {\ensuremath{\overline \rho}\xspace}
\def\etabar {\ensuremath{\overline \eta}\xspace}
\def\meas {\ensuremath{|V_{cb}|, |\frac{V_{ub}}{V_{cb}}|, |\varepsilon_K|, \Delta m_{B_d}}\xspace}

\def\DeltaS     {\mbox{$\Delta S$}\xspace}
\def\DeltaC     {\mbox{$\Delta C$}\xspace}

\def\epstag  {\ensuremath{\varepsilon_{\rm tag}}\xspace}
\def\epstagbz  {\ensuremath{\varepsilon_{\Bz}}\xspace}
\def\epstagbzb  {\ensuremath{\varepsilon_{\Bzb}}\xspace}
\def\epstagc  {\ensuremath{\varepsilon_{\rm c}}\xspace}
\def\epstagasym  {\ensuremath{\Delta\varepsilon_{\rm tag}}\xspace}
\def\wtag    {\ensuremath{w}\xspace}
\def\wtagbz    {\ensuremath{w_{\Bz}}\xspace}
\def\wtagbzb   {\ensuremath{w_{\Bzb}}\xspace}
\def\wtagc    {\ensuremath{w_{\rm c}}\xspace}
\def\wtagasym    {\ensuremath{\Delta w}\xspace}
\def\tagfac  {\ensuremath{\epstag(1-2\wtag)^2}\xspace}
\def\tagfacc  {\ensuremath{\epstagc(1-2\wtagc)^2}\xspace}
\def\qtag    {\ensuremath{Q}\xspace}

\def\tagvarew {\ensuremath{E^{W}_{90}}\xspace}

\def\bbrleptontag{{\tt Lepton}\xspace}
\def\bbrkaonitag{{\tt Kaon\,I}\xspace}
\def\bbrkaoniitag{{\tt Kaon\,II}\xspace}
\def\bbrkpitag{{\tt Kaon-Pion}\xspace}
\def\bbrpiontag{{\tt Pion}\xspace}
\def\bbrothertag{{\tt Other}\xspace}
\def\bbruntag{{\tt Untagged}\xspace}

%
%
\newcommand\vud {\ensuremath{V_{\mathrm{ud}}}\xspace}
\newcommand\vus {\ensuremath{V_{\mathrm{us}}}\xspace}
\newcommand\vub {\ensuremath{V_{\mathrm{ub}}}\xspace}
\newcommand\vcd {\ensuremath{V_{\mathrm{cd}}}\xspace}
\newcommand\vcs {\ensuremath{V_{\mathrm{cs}}}\xspace}
\newcommand\vcb {\ensuremath{V_{\mathrm{cb}}}\xspace}
\newcommand\vtd {\ensuremath{V_{\mathrm{td}}}\xspace}
\newcommand\vts {\ensuremath{V_{\mathrm{ts}}}\xspace}
\newcommand\vtb {\ensuremath{V_{\mathrm{tb}}}\xspace}
\def\vckm       {\ensuremath{{V}_{\rm CKM}}\xspace}
\def\theckmmatrix  {\ensuremath{ \left( \begin{array}{ccc} \vud & \vus & \vub \\ \vcd & \vcs & \vcb \\ \vtd & \vts & \vtb \end{array}\right).}\xspace}

\def\Vud  {\ensuremath{|V_{ud}|}\xspace}
\def\Vcd  {\ensuremath{|V_{cd}|}\xspace}
\def\Vtd  {\ensuremath{|V_{td}|}\xspace}
\def\Vus  {\ensuremath{|V_{us}|}\xspace}
\def\Vcs  {\ensuremath{|V_{cs}|}\xspace}
\def\Vts  {\ensuremath{|V_{ts}|}\xspace}
\def\Vub  {\ensuremath{|V_{ub}|}\xspace}
\def\Vcb  {\ensuremath{|V_{cb}|}\xspace}
\def\Vtb  {\ensuremath{|V_{tb}|}\xspace}

\newcommand\phione   {\ensuremath{\phi_1}\xspace}
\newcommand\phitwo   {\ensuremath{\phi_2}\xspace}
\newcommand\phithree {\ensuremath{\phi_3}\xspace}


\def\stwoa    {\ensuremath{\sin\! 2 \alpha  }\xspace}
\def\stwob    {\ensuremath{\sin\! 2 \beta   }\xspace}
\def\stwog    {\ensuremath{\sin\! 2 \gamma  }\xspace}
\def\mistag   {\ensuremath{w}\xspace}
\def\dilution {\ensuremath{\cal D}\xspace}
\def\deltaz   {\ensuremath{{\rm \Delta}z}\xspace}
\def\deltat   {\ensuremath{{\rm \Delta}t}\xspace}
\def\deltamd  {\ensuremath{{\rm \Delta}m_d}\xspace}
\def\deltaG   {\ensuremath{{\rm \Delta}\Gamma}\xspace}

\def\clong   {\ensuremath{C_{long}}}
\def\slong   {\ensuremath{S_{long}}}
\def\ct      {\ensuremath{C_{tran}}}
\def\st      {\ensuremath{S_{tran}}}
\def\fL      {\ensuremath{f_{L}}}

\def\Acp     {\ensuremath{A_{CP} }}

\def\Btopp {\ensuremath{B \rightarrow PP}}
\def\Btovv {\ensuremath{B \rightarrow VV}}
\def\Btopv {\ensuremath{B \rightarrow PV}}
\def\Btotv {\ensuremath{B \rightarrow TV}}

\newcommand{\fsubd}{\ensuremath{f_D}}\xspace
\newcommand{\fds}{\ensuremath{f_{D_s}}\xspace}
\newcommand{\fsubb}{\ensuremath{f_B}\xspace}
\newcommand{\fbd}{\ensuremath{f_{B_d}}\xspace}
\newcommand{\fbs}{\ensuremath{f_{B_s}}\xspace}
\newcommand{\bsubb}{\ensuremath{B_B}\xspace}
\newcommand{\bbd}{\ensuremath{B_{B_d}}\xspace}
\newcommand{\bbs}{\ensuremath{B_{B_s}}\xspace}
\newcommand{\rgbb}{\ensuremath{\hat{B}_B}\xspace}
\newcommand{\rgbbd}{\ensuremath{\hat{B}_{B_d}}\xspace}
\newcommand{\rgbbs}{\ensuremath{\hat{B}_{B_s}}\xspace}
\newcommand{\rgbk}{\ensuremath{\hat{B}_K}\xspace}
\newcommand{\lqcd}{\ensuremath{\Lambda_{\mathrm{QCD}}}\xspace}

\newcommand{\su}     [1]  {\ensuremath{SU(#1)}}
\newcommand{\e}      [1]   { {\ensuremath{ \times 10^{ {#1} } }}}

\newcommand{\secref}[1]{Section~\ref{sec:#1}}
\newcommand{\subsecref}[1]{Section~\ref{subsec:#1}}
\newcommand{\figref}[1]{Figure~\ref{fig:#1}}
\newcommand{\tabref}[1]{Table~\ref{tab:#1}}


\newcommand{\epjBase}        {Eur.\ Phys.\ Jour.\xspace}
\newcommand{\jprlBase}       {Phys.\ Rev.\ Lett.\xspace}
\newcommand{\jprBase}        {Phys.\ Rev.\xspace}
\newcommand{\jplBase}        {Phys.\ Lett.\xspace}
\newcommand{\nimBaseA}       {Nucl.\ Instrum.\ Methods Phys.\ Res., Sect.\ A\xspace}
\newcommand{\nimBaseB}       {Nucl.\ Instrum.\ Methods Phys.\ Res., Sect.\ B\xspace}
\newcommand{\nimBaseC}       {Nucl.\ Instrum.\ Methods Phys.\ Res., Sect.\ C\xspace}
\newcommand{\nimBaseD}       {Nucl.\ Instrum.\ Methods Phys.\ Res., Sect.\ D\xspace}
\newcommand{\nimBaseE}       {Nucl.\ Instrum.\ Methods Phys.\ Res., Sect.\ E\xspace}
\newcommand{\npBase}         {Nucl.\ Phys.\xspace}
\newcommand{\zpBase}         {Z.\ Phys.\xspace}

\newcommand{\apas}      [1]  {{Acta Phys.\ Austr.\ Suppl.\ {\bf #1}}}
\newcommand{\app}       [1]  {{Acta Phys.\ Polon.\ {\bf #1}}}
\newcommand{\ace}       [1]  {{Adv.\ Cry.\ Eng.\ {\bf #1}}}
\newcommand{\anp}       [1]  {{Adv.\ Nucl.\ Phys.\ {\bf #1}}}
\newcommand{\annp}      [1]  {{Ann.\ Phys.\ {\bf #1}}}
\newcommand{\araa}      [1]  {{Ann.\ Rev.\ Astr.\ Ap.\ {\bf #1}}}
\newcommand{\arnps}     [1]  {{Ann.\ Rev.\ Nucl.\ Part.\ Sci.\ {\bf #1}}}
\newcommand{\arns}      [1]  {{Ann.\ Rev.\ Nucl.\ Sci.\ {\bf #1}}}
\newcommand{\appopt}    [1]  {{Appl.\ Opt.\ {\bf #1}}}
\newcommand{\japj}      [1]  {{Astro.\ Phys.\ J.\ {\bf #1}}}
\newcommand{\baps}      [1]  {{Bull.\ Am.\ Phys.\ Soc.\ {\bf #1}}}
\newcommand{\seis}      [1]  {{Bull.\ Seismological Soc.\ of Am.\ {\bf #1}}}
\newcommand{\cmp}       [1]  {{Commun.\ Math.\ Phys.\ {\bf #1}}}
\newcommand{\cnpp}      [1]  {{Comm.\ Nucl.\ Part.\ Phys.\ {\bf #1}}}
\newcommand{\cpc}       [1]  {{Comput.\ Phys.\ Commun.\ {\bf #1}}}
\newcommand{\epj}       [1]  {\epjBase\ {\bf #1}}
\newcommand{\epjc}      [1]  {\epjBase\ C~{\bf #1}}
\newcommand{\fizika}    [1]  {{Fizika~{\bf #1}}}
\newcommand{\fp}        [1]  {{Fortschr.\ Phys.\ {\bf #1}}}
\newcommand{\ited}      [1]  {{IEEE Trans.\ Electron.\ Devices~{\bf #1}}}
\newcommand{\itns}      [1]  {{IEEE Trans.\ Nucl.\ Sci.\ {\bf #1}}}
\newcommand{\ijqe}      [1]  {{IEEE J.\ Quantum Electron.\ {\bf #1}}}
\newcommand{\ijmp}      [1]  {{Int.\ Jour.\ Mod.\ Phys.\ {\bf #1}}}
\newcommand{\ijmpa}     [1]  {{Int.\ J.\ Mod.\ Phys.\ {\bf A{\bf #1}}}}
\newcommand{\jl}        [1]  {{JETP Lett.\ {\bf #1}}}
\newcommand{\jetp}      [1]  {{JETP~{\bf #1}}}
\newcommand{\jpg}       [1]  {{J.\ Phys.\ {\bf G{\bf #1}}}}
\newcommand{\jap}       [1]  {{J.\ Appl.\ Phys.\ {\bf #1}}}
\newcommand{\jmp}       [1]  {{J.\ Math.\ Phys.\ {\bf #1}}}
\newcommand{\jmes}      [1]  {{J.\ Micro.\ Elec.\ Sys.\ {\bf #1}}}
\newcommand{\mpl}       [1]  {{Mod.\ Phys.\ Lett.\ {\bf #1}}}

\newcommand{\nim}       [1]  {\nimBaseC~{\bf #1}}
\newcommand{\nimaa}     [1]  {\nimBaseE~{\bf #1}}
\newcommand{\nima}      [1]  {\nimBaseA~{\bf #1}}
\newcommand{\nimb}      [1]  {\nimBaseB~{\bf #1}}
\newcommand{\nimc}      [1]  {\nimBaseC~{\bf #1}}
\newcommand{\nimd}      [1]  {\nimBaseD~{\bf #1}}
\newcommand{\nime}      [1]  {\nimBaseE~{\bf #1}}
\newcommand{\oldnim}    [1]  {Nucl.\ Instrum.\ Methods Phys.\ Res.~{\bf #1}}
\newcommand{\oldernim}  [1]  {Nucl.\ Instrum.\ Methods~{\bf #1}}

\newcommand{\np}        [1]  {\npBase\ {\bf #1}}
\newcommand{\npb}       [1]  {\npBase\ B~{\bf #1}}
\newcommand{\npps}      [1]  {{Nucl.\ Phys.\ Proc.\ Suppl.\ {\bf #1}}}
\newcommand{\npaps}     [1]  {{Nucl.\ Phys.\ A~Proc.\ Suppl.\ {\bf #1}}}
\newcommand{\npbps}     [1]  {{Nucl.\ Phys.\ B~Proc.\ Suppl.\ {\bf #1}}}

\newcommand{\ncim}      [1]  {{Nuo.\ Cim.\ {\bf #1}}}
\newcommand{\optl}      [1]  {{Opt.\ Lett.\ {\bf #1}}}
\newcommand{\optcom}    [1]  {{Opt.\ Commun.\ {\bf #1}}}
\newcommand{\partacc}   [1]  {{Particle Acclerators~{\bf #1}}}
\newcommand{\pan}       [1]  {{Phys.\ Atom.\ Nuclei~{\bf #1}}}
\newcommand{\pflu}      [1]  {{Physics of Fluids~{\bf #1}}}
\newcommand{\ptoday}    [1]  {{Physics Today~{\bf #1}}}

\newcommand{\jpl}       [1]  {\jplBase\ {\bf #1}}
\newcommand{\plb}       [1]  {\jplBase\ B~{\bf #1}}
\newcommand{\prep}      [1]  {{Phys.\ Rep.\ {\bf #1}}}
\newcommand{\jprl}      [1]  {\jprlBase\ {\bf #1}}
\newcommand{\pr}        [1]  {\jprBase\ {\bf #1}}
\newcommand{\jpra}      [1]  {\jprBase\ A~{\bf #1}}
\newcommand{\jprd}      [1]  {\jprBase\ D~{\bf #1}}
\newcommand{\jpre}      [1]  {\jprBase\ E~{\bf #1}}

\newcommand{\prsl}      [1]  {{Proc.\ Roy.\ Soc.\ Lond.\ {\bf #1}}}
\newcommand{\ppnp}      [1]  {{Prog.\ Part.\ Nucl.\ Phys.\ {\bf #1}}}
\newcommand{\progtp}    [1]  {{Prog.\ Theor.\ Phys.\ {\bf #1}}}
\newcommand{\rpp}       [1]  {{Rep.\ Prog.\ Phys.\ {\bf #1}}}
\newcommand{\jrmp}      [1]  {{Rev.\ Mod.\ Phys.\ {\bf #1}}}  
\newcommand{\rsi}       [1]  {{Rev.\ Sci.\ Instr.\ {\bf #1}}}
\newcommand{\sci}       [1]  {{Science~{\bf #1}}}
\newcommand{\sjnp}      [1]  {{Sov.\ J.\ Nucl.\ Phys.\ {\bf #1}}}
\newcommand{\spd}       [1]  {{Sov.\ Phys.\ Dokl.\ {\bf #1}}}
\newcommand{\spu}       [1]  {{Sov.\ Phys.\ Usp.\ {\bf #1}}}
\newcommand{\tmf}       [1]  {{Teor.\ Mat.\ Fiz.\ {\bf #1}}}
\newcommand{\yf}        [1]  {{Yad.\ Fiz.\ {\bf #1}}}
\newcommand{\zp}        [1]  {\zpBase\ {\bf #1}}
\newcommand{\zpc}       [1]  {\zpBase\ C~{\bf #1}}
\newcommand{\zpr}       [1]  {{ZhETF Pis.\ Red.\ {\bf #1}}}

\newcommand{\hepex}     [1]  {hep-ex/{#1}}
\newcommand{\hepph}     [1]  {hep-ph/{#1}}
\newcommand{\hepth}     [1]  {hep-th/{#1}}


\def\aslund     {\mbox{\tt Aslund}\xspace}
\def\bbsim      {\mbox{\tt BBsim}\xspace}
\def\beast      {\mbox{\tt Beast}\xspace}
\def\beget      {\mbox{\tt Beget}\xspace}
\def\Bta        {\mbox{\tt Beta}\xspace}
\def\betakfit   {\mbox{\tt BetaKfit}\xspace}
\def\cornelius  {\mbox{\tt Cornelius}\xspace}
\def\evtgen     {\mbox{\tt EvtGen}\xspace}
\def\euclid     {\mbox{\tt Euclid}\xspace}
\def\fitver     {\mbox{\tt FitVer}\xspace}
\def\fluka      {\mbox{\tt Fluka}\xspace}
\def\fortran    {\mbox{\tt Fortran}\xspace}
\def\gcalor     {\mbox{\tt GCalor}\xspace}
\def\geant      {\mbox{\tt GEANT}\xspace}
\def\gheisha    {\mbox{\tt Gheisha}\xspace}
\def\hemicosm   {\mbox{\tt HemiCosm}\xspace}
\def\hepevt     {\mbox{\tt{/HepEvt/}}\xspace}
\def\jetset74   {\mbox{\tt Jetset \hspace{-0.5em}7.\hspace{-0.2em}4}\xspace}
\def\koralb     {\mbox{\tt KoralB}\xspace}
\def\minuit     {\mbox{\tt Minuit}\xspace}
\def\objegs     {\mbox{\tt Objegs}\xspace}
\def\paw        {\mbox{\tt Paw}\xspace}
\def\Root       {\mbox{\tt Root}\xspace}
\def\squaw      {\mbox{\tt Squaw}\xspace}
\def\stdhep     {\mbox{\tt StdHep}\xspace}
\def\trackerr   {\mbox{\tt TrackErr}\xspace}
\def\turtle     {\mbox{\tt Decay~Turtle}\xspace}

\def\babar{\mbox{\slshape B\kern-0.1em{\smaller A}\kern-0.1em B\kern-0.1em{\smaller A\kern-0.2em R}}\xspace}
\def\belle{\mbox{\normalfont Belle}\xspace}
\def\superb {\ensuremath{\mbox{Super$B$}}\xspace}

%

\contribution[The Physics Potential of SuperB]  
{The Physics Potential of SuperB}  
{F. F.}{Wilson}  
{STFC Rutherford Appleton Laboratory, Chilton, Didcot, Oxon, OX11 0QX, UK}  
{Fergus.Wilson@stfc.ac.uk}  
{on behalf of the SuperB Collaboration}  
%

\abstract{%
\superb\ is a major new European \epem\ collider facility to
  be built in Italy that will provide a precise study of the 
  structure of New Physics beyond the
  Standard Model at energy scales above the LHC as well as a
  comprehensive program of Standard Model physics. In this article, I
  review the physics opportunities, the status of the accelerator and
  detector studies, and the future plans. 
}
%

\section{Introduction}


The new \superb\ facility will investigate the consequences for
flavour physics of any discoveries at the LHC and search for
New Physics (NP) signatures at energy scales that exceed the direct
search capabilities of the LHC. A super-flavour factory will also be
able to improve the precision and sensitivity of the previous
generation of flavour factories by factors of five to ten. The sides
and angles of the Unitarity Triangle will be determined to an accuracy
of $\sim 1\%$. Limits on Lepton Flavour Violation (LFV) in $\tau$
decays will be improved by two orders of magnitude. It will become
feasible to search for CP violation (CPV) in charm mixing. Extensive
searches for new states in bottomium and charmonium spectroscopy will
be achieved.  New precision measurements of electroweak properties,
such as the running of the weak mixing angle $\sin^2\theta_W$ with
energy, should become possible.

Flavour physics is an ideal tool for indirect searches for NP. Both
mixing and CPV in \B\ and \D\ mesons occur at the loop level in
the Standard Model (SM) and therefore can be subject to NP
corrections. New virtual particles occurring in the loops or tree
diagrams can also change the predicted branching fractions or angular
distributions of rare decays. Current experimental limits indicate NP
with trivial flavour couplings has a scale in the 10-100\tev\ range,
which is much higher than the 1 \tev\ scale suggested by SM Higgs
physics. This means that either
the NP scale can not be seen in direct searches at the LHC or
the NP scale is close to 1\tev\ and therefore the flavour structure of
the NP must be very complex. In either case, indirect searches provide
a way of understanding the new phenomena in great detail.

\superb\ is an asymmetric \epem\ collider with a 1.3\km\
circumference. The design calls for 6.7\gev\ positrons colliding with
4.18\gev electrons at a centre of mass energy $\sqrt{s} =
10.58\gev$. The boost $\beta\gamma = 0.238$ is approximately half the
value used at \babar~\cite{ref:babar}. The electron beam can be
60\%-80\% polarized. The design luminosity is
$10^{36}$\cms and data taking is expected to start in the latter part
of this decade with a delivered integrated luminosity of 75\invab\
over five years. It should be possible to exceed the baseline luminosity
specification, leading to the prospect of collecting
20-40\invab\ per year in later years.

In the following sections, I discuss the physics potential of some of
the key measurements to be made at the \superb\ factory with an
integrated luminosity of 75 \invab. In addition, there is a
comprehensive program for \Bs\ at the \FiveS\ resonance, bottomium and
charmonium spectroscopy, ISR physics, g-2 hadronic contributions, and two-photon
interactions, to name just a few.

\section{Physics Potential}

Both \babar\ and \belle~\cite{ref:belle} have successfully measured
the CKM Unitarity Triangle angles $\alpha$, $\beta$ and
$\gamma$~\cite{ref:hfag}. Although there are discrepancies in some
measurements, overall everything is consistent to a few
sigma. Increasing the statistics will show if these tensions are real
and possible signs of NP. It will be possible to measure the angles
$\alpha$ and $\gamma$ to $1-2\%$, and $\beta$ to 0.1\%. \Vcb\ and
\Vub\ can be measured to 1\% and 2\% accuracy, respectively, in both
inclusive and exclusive semileptonic decays. The production of copious
amounts of charm decays could lead to the measurement of the charm
Unitarity Triangle parameters.  Figure~\ref{fig:ckm} shows the
\rhobar-\etabar\ plane with current and predicted experimental
measurements, assuming the current measurements maintain their central
values.

\begin{figure}[htb!]
\begin{center}
  \includegraphics[height=7cm]{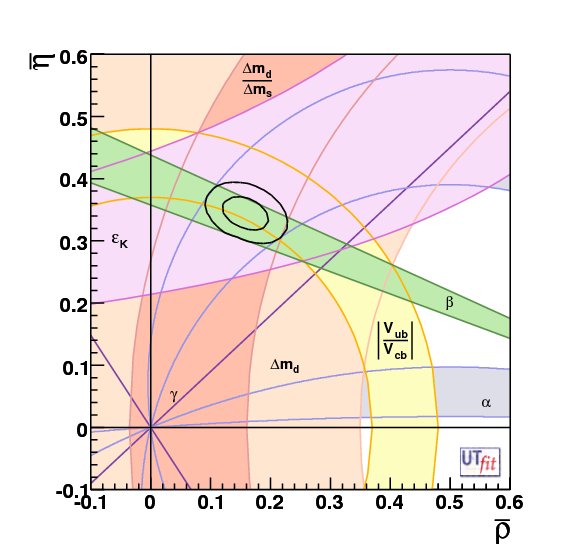}
  \includegraphics[height=7cm]{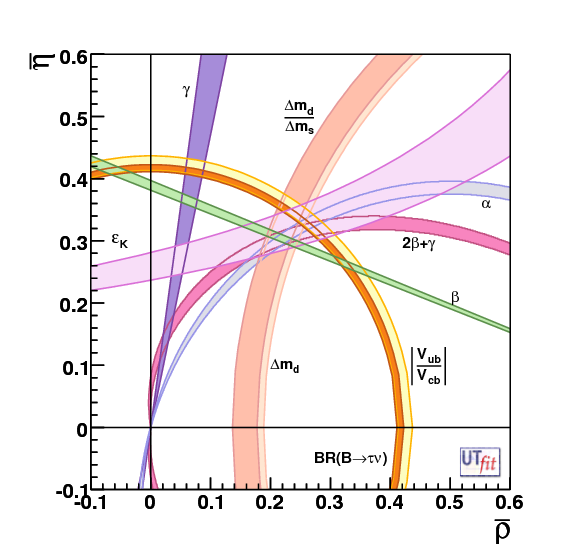}
\end{center}
\caption{Regions corresponding to 95\% probability for \rhobar\ and
  \etabar\ with current measurements (left) and with \superb\
  precision assuming the current central values (right).}
\label{fig:ckm}
\end{figure}

\superb\ will make precision measurements of a series of ``Golden
Modes''. The SM predictions for these modes are well calculated and
they can be cleanly measured experimentally. NP scenarios
can be differentiated by comparing the measured values with NP
predictions. Table~\ref{tab:golden} shows just some of the key
measurements and a sample of NP models.

\begin{table}
\begin{center}
\resizebox{1.0\textwidth}{!}{
\begin{tabular}{l|cccccc|ccccc}
\hline
   & $H^+$           & MFV & non-MFV & NP         & Right-hand & LTH &
   \multicolumn{5}{c}{SUSY models}  \\
   & high tan$\beta$ &     &         & Z-penguins & currents   &     &
   AC &RVV2 &AKM &$\delta LL$ & FBMSSM\\
\hline
$\calB(\tau\to\mu\gamma)$  & &   &   & &   &   & L & L & M & L & L\\
$\calB(\tau\to\mu\mu\mu)$  & &   &   & &   & L & \\
\hline
$\calB(\B\to \tau\nu,\mu\nu)$     & L-CKM & &   & &   & &  \\
$\calB(\B\to K^{(*)}\nu\nub)$    & &   & M & L  & & & M & M & M & M & M \\
$S_{\KS\piz\gamma}$        & &   &   & & L  & &  \\
Angle $\beta$ ($\Delta S$) & &   & L-CKM & & L & & L & M & M & L & L \\
$\Acp(\B\to X_s\gamma)$    & &   & L & & M & & M & M & M & L & L  \\
$\calB(\B\to X_s\gamma)$   & & L & M & & M & & \\
$\calB(\B\to X_sll)$       & &   & M & M & M & &  \\
$A_{FB}(\B\to K^{(*)}ll)$  & &   &  &  &  & & M & M & M & L & L  \\
\hline
Charm mixing               &   &  &   & &  & & L & M & M & M & M \\ 
CPV in  Charm              & L &  &   & &  & &   &   &   & L & \\ 
\hline
\end{tabular}
}
\end{center}
\caption{The golden matrix of observables versus a sample of NP
  scenarios. MFV is a representative Minimal Flavour Violation model;
  LTH is a Littlest Higgs Model with T Parity. A number of explicit SUSY
  models are included~\cite{ref:physics}. L denotes a large
  effect, M a measurable effect and L-CKM indicates a measurement that
  requires precise measurement of the CKM matrix. $\Delta S$ is the
  difference in the angle $\beta$ between $b\to\s$ penguin-dominated
  transitions and $b\to\ccbar\s$ decays.}
\label{tab:golden}
\end{table}

In 2-Higgs-doublet (2HDM-II) and MSSM models, the decay $\B\to\tau\nu$
is sensitive to the presence of a charged Higgs $H^-$ replacing the SM
\Wm. \superb\ will be able to exclude masses up to $\sim 2-3$\tev\ for
values of $\tan\beta$ up to 80. The region of charged Higgs mass
versus $\tan\beta$ that can be excluded is shown in
Figure~\ref{fig:hplus} for both the 2HDM-II and MSSM models. This
includes the current 20\% uncertainty from $f_b$ and \vub\ that can be
expected to be much reduced in the future.

\begin{figure}[htb!]
\begin{center}
  \includegraphics[height=6cm]{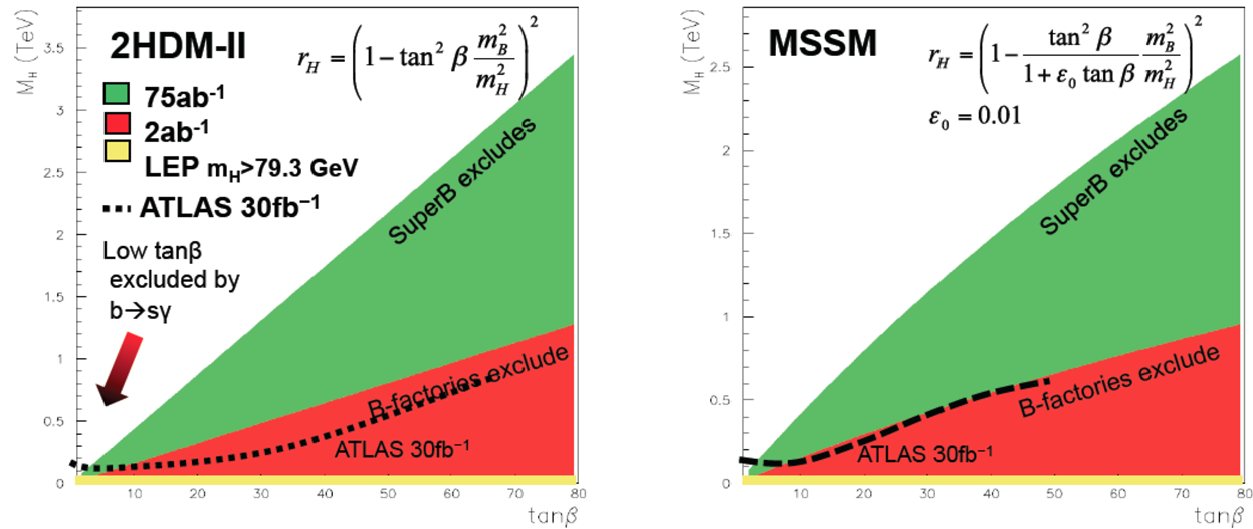}
\end{center}
\caption{The mass of the charged Higgs versus $\tan\beta$ from
  $\B\to\tau\nu$ decays for a 2HDM-II (left) and MSSM (right)
  model. The dark (red) region is excluded assuming the \babar\
  and \belle\ datasets are combined and the light (green) region shows the exclusion
  potential of \superb.}
\label{fig:hplus}
\end{figure}

\superb\ can access the off-diagonal elements of generic squark mass
matrices in the MSSM model using the mass insertion
approximation. These can not be seen by the LHC general purpose
detectors. As an example, \superb\ is sensitive to non-zero values of
the matrix element $(\delta^d_{23})_{LL,LR}$ for gluino masses in the
1-10\tev\ range through decays such as $b\to s\gamma$ and $b\to
sl^+l^-$ (Figure~\ref{fig:mssm}).

\begin{figure}[htb!]
\begin{center}
  \includegraphics[width=7cm]{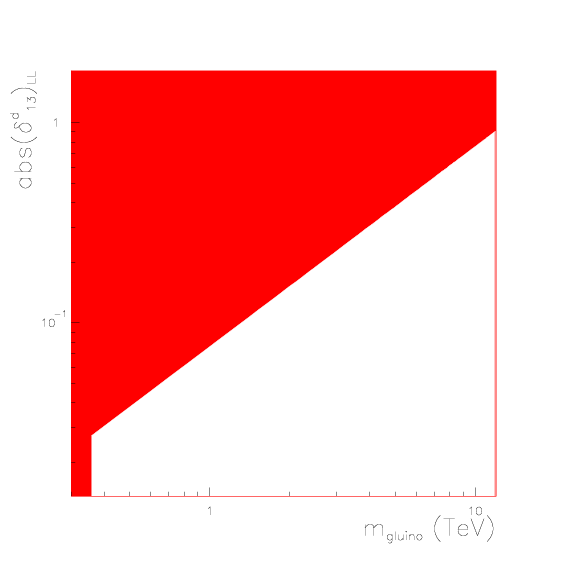}
  \includegraphics[width=7cm]{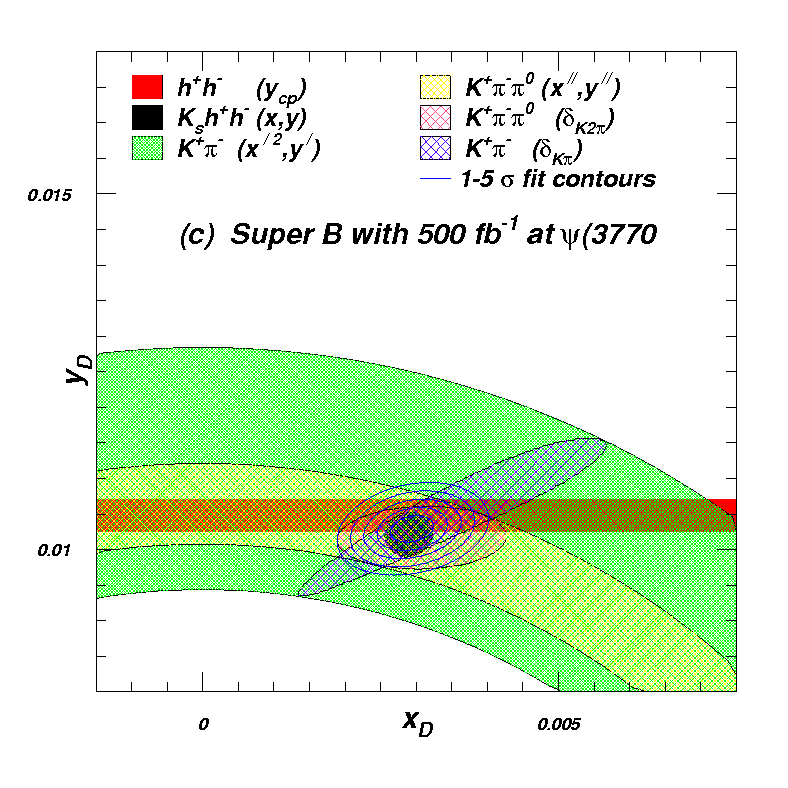}
\end{center}
\caption{Left: The shaded (red) region shows where a measurement can be made
  (defined as a 3$\sigma$ significance) of the matrix element
  $(\delta^d_{23})_{LL,LR}$ as a function of gluino mass in an MSSM
  model from measurements involving a $b\to s$ transition. Right: the
  expected precision on charm mixing parameters from combining BES-III and
  \superb\ $\psi(3770)$ and \FourS\ data.}
\label{fig:mssm}
\end{figure}


An almost equal number of $\taup\taum$ pairs are produced as \BB\
pairs at the \FourS\ resonance. Current experimental 90\% confidence
level upper limits on $\tau$ LFV are in the $10^{-8}-10^{-7}$ range, 
depending on the decay. In the very clean environment of \superb,
upper limits on $\tau$ LFV can be achieved down to a level of $2\times
10^{-10}$ for $\tau\to\mu\mu\mu$ and \superb\ can measure the upper
limits in $\sim 50$  other $\tau$ decay modes.
Background-free modes should scale with the luminosity while 
other modes will scale with $\sqrt{\calL}$ or better, thanks to
re-optimized analysis techniques.  In $\tau\to\mu\gamma$ for example,
LFV is predicted at the level $10^{-10}-10^{-7}$ depending on the NP
model. SU(5) SUSY GUT models predict  $\tau\to\mu\gamma$ branching fractions between
$10^{-7}$ and $10^{-9}$ depending on the NP phase, so the majority of the
parameter space is within the expected \superb\ sensitivity of
$2\times 10^{-9}$. 

\begin{figure}[htb!]
\begin{center}
  \includegraphics[width=0.50\textwidth]{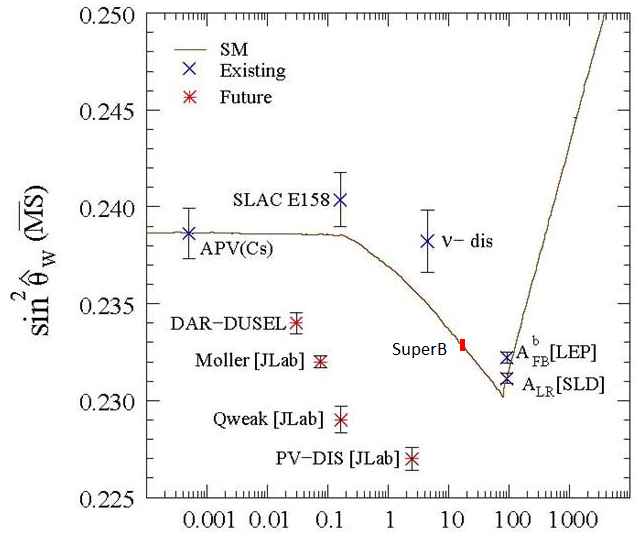}
  \includegraphics[width=0.45\textwidth]{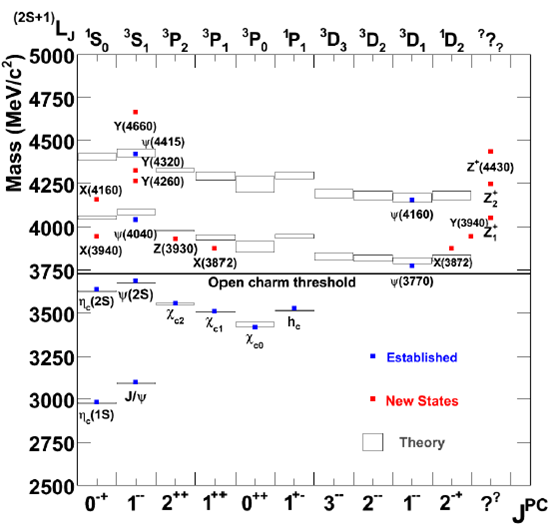}
\end{center}
\caption{Left: Measurements of $\sin^2\theta_w$ as a function of
  energy (\gev). The size of the bar at an energy $\sim10.6\gev$
  representing the \superb\ measurement is approximately the same size
  as the error. Right: Measured masses of newly observed states
  positioned according to their most likely quantum numbers.}
\label{fig:lfv2}
\end{figure}


CPV in charm decays is expected to be very low in the SM ($<1\%$) so
its detection would be a clear indicator of NP. Current values for the
mixing parameters $x$ and $y$ from HFAG~\cite{ref:hfag} fits give
$(0.63\pm0.20)$\% and $(0.75\pm0.12)$\%, respectively, allowing for
CPV~\cite{ref:pdg}. At \superb, the errors should reduce to $0.07$\%
and $0.02$\%, respectively. If the results are combined with expected
results from BES-III and a dedicated \superb\ 500\invfb run ($\sim$ 4
months running) at the \D\Db\ threshold, the BES-III/CLEO-c physics programme
can be repeated leading to a further reduction in these errors to
$0.02$\% and $0.01$\%, respectively. This is shown in the right-hand
plot of Figure~\ref{fig:mssm}.


If a polarised electron beam is available, many of the upper limits on
$\tau$ LFV modes can be improved by an additional factor of two. The
polarisation also allows for the search for $\tau$ EDM at a level of
$2\times 10^{-19} e\cm$ and measurement of the $\tau$ anomalous magnetic
moment $\Delta\alpha_{\tau}$ with
an error of $10^{-6}$. The value of $\sin^2\theta_w$ can be measured
with an accuracy $\pm 1.8\times 10^{-4}$ at $Q=10.58\gev$ and so help
understand the discrepancy in the measurements from LEP, SLD and
NuTev~\cite{ref:ew}. This is shown in the left-hand plot of
Figure~\ref{fig:lfv2} where the size of the bar at $Q=10.58\gev$
represents the expected error on the \superb\ measurement. It may
even be possible to measure $\sin^2\theta_w$ at the $\psi(3770)$ mass
if polarisation can be achieved.


The B-Factories and the Tevatron have discovered heavy bound states
that do not fit into the conventional meson interpretation. However,
apart from some exceptions like the $X(3872)$, they have only been
observed in a single decay channel with a significance only just above
$5\sigma$. The right-hand plot of Figure~\ref{fig:lfv2} shows some of
the newly discovered states. Possible explanations include hybrids,
molecules, tetraquarks and threshold effects. \superb's ability to run
at the $\Upsilon(nS)$ resonances and charm threshold provides a unique
opportunity for testing low- and high-energy QCD
predictions. Predicting the expected rates for poorly measured
resonances is of course hard and work is on-going to improve the
extrapolations. The $\B\to X(3872)K$ decays should produce $\sim
2k-10k$ events in each of their main decay
channels. $Y(4260)\to\jpsi\pip\pim$ will have $\sim45k$ events, while
$\sim4.5k$ events can be expected for both $Y(4350)$ and $Y(4660)$
decaying to $\psi(2S)\pip\pim$. It should be possible to confirm the
existence of the $Z_1^+(4050)$, $Z^+(4430)$ and $Z_2^+(4430)$ as \superb\ will
collect between $150k - 2M$ events of the relevant fully reconstructed
final states $\jpsi\pip K$, $\psi(2S)\pip K$, and $\chi_{cJ}\pip K$.

\section{Status of the project}

The physics potential~\cite{ref:physics}, and the
detector~\cite{ref:detector} and accelerator~\cite{ref:accelerator}
plans have been extensively documented.  The accelerator parameters
are close to final for operating in the $\psi(3770)$ to \FiveS\ energy
range and the accelerator will reuse large parts of the SLAC PEP-II
hardware. The campus of Tor Vergata University, Rome, was chosen as
the site at the end of May 2011. Data taking should begin five to six
years after construction begins.

}  


%% file: hadron2011.bbl
\begin{thebibliography}{99}
\bibitem{ref:babar} 
B.~Aubert et al., (\babar\ Collaboration), \nima{479}, 1 (2002).


\bibitem{ref:belle} 
A.~Abashian et al., (\belle\ Collaboration), \nima{479}, 1 (2002).
 
\bibitem{ref:hfag} 
Heavy Flavor Averaging Group (HFAG), www.slac.stanford.edu/xorg/hfag.

\bibitem{ref:pdg} 
C.~Amsler et al., \jpg{37}, (2010) 075021.

\bibitem{ref:ew}
EW Working Groups, \emph{Precision Electroweak measurements on the Z Resonance},
  Phys. Rept. {\bf 427}, 257 (2006).

\bibitem{ref:physics} 
D.G.~Hitlin et al., \emph{New Physics at the Super Flavor Factory}, [arXiv:0810.1312.1541];
M.~Bona et al., \emph{\superb Conceptual Design Report}, [arXiv:0709.0451];
 B.~O'Leary et al., \emph{\superb Progress Report -- Physics}, [arXiv:1008.1541].

\bibitem{ref:detector} 
E.~Grauges et al., \emph{\superb Progress Report -- Detector}, [arXiv:1007.4241].

\bibitem{ref:accelerator} 
M.E.~Biagini et al., \emph{\superb Progress Report -- Accelerator}, [arXiv:1009.6178].

\end{thebibliography}
